\documentclass{elsarticle}

\usepackage{bm}
\usepackage{graphicx}
\usepackage{color}
\usepackage{tabularx}
\usepackage{amsmath}
\usepackage{amsfonts}
\usepackage{amssymb}

\usepackage{algpseudocode}
\usepackage{algorithm}

\usepackage{lineno}
\usepackage[colorlinks=true,pdftex]{hyperref}

\modulolinenumbers[5]

\journal{Computer Physics Communications}

\newcommand{\Tr}{ {\rm Tr} \, }
\newcommand{\tr}{ {\rm Tr} \, }

\newcommand{\lr}[1]{ \left( #1 \right) }
\newcommand{\lrs}[1]{ \left[ #1 \right] }

\newcommand{\ket}[1]{\ensuremath{\left|#1\right\rangle}}
\newcommand{\bra}[1]{\ensuremath{\left\langle#1\right|}}

\graphicspath{
	{./}
	{./figures/}
}

\begin{document}

\begin{frontmatter}

\title{Schur complement solver for Quantum Monte-Carlo simulations of strongly interacting fermions}

\author[AD1]{Maksim~Ulybyshev\corref{mycorrespondingauthor}}
\cortext[mycorrespondingauthor]{Corresponding author.}
\ead{maksim.ulybyshev@physik.uni-wuerzburg.de}

\author[AD2]{Nils~Kintscher}

\author[AD2]{Karsten~Kahl}

\author[AD3]{Pavel~Buividovich}

\address[AD1]{Institute of Theoretical Physics, W\"urzburg University, 97074 W\"urzburg, Germany}
\address[AD2]{School of Mathematics and Natural Sciences, University of Wuppertal, 42097 Wuppertal, Germany}
\address[AD3]{Institute of Theoretical Physics, Regensburg University, 93040 Regensburg, Germany}

\begin{abstract}
We present a non-iterative solver based on the Schur complement method for sparse linear systems of special form which appear in Quantum Monte-Carlo (QMC) simulations of strongly interacting fermions on the lattice. While the number of floating-point operations for this solver scales as the cube of the number of lattice sites, for practically relevant lattice sizes it is still significantly faster than iterative solvers such as the Conjugate Gradient method in the regime of strong inter-fermion interactions, for example, in the vicinity of quantum phase transitions. The speed-up is even more dramatic for the solution of multiple linear systems with different right-hand sides. We present benchmark results for QMC simulations of the tight-binding models on the hexagonal graphene lattice with on-site (Hubbard) and non-local (Coulomb) interactions, and demonstrate the potential for further speed-up using GPU. 
\end{abstract}


\end{frontmatter}


\section{Introduction}
\label{sec:introduction}

Quantum Monte Carlo (QMC) is one of the most powerful numerical techniques for studying strongly-correlated many-body quantum systems. QMC is widely used both in high-energy physics, with lattice Quantum Chromodynamics (QCD) being probably the most important application, and in condensed matter physics, where we will refer mostly to determinantal QMC.

For models with fermionic degrees of freedom, such as lattice QCD or tight-binding models which describe electronic band structure in solids, the so-called fermionic determinant $\det\lr{M}$ appears in the path integral representation of quantum expectation values. This determinant involves a large sparse matrix $M$ of some special structure, and its treatment is usually the most computationally intensive and algorithmically complex part of QMC algorithms.

A popular QMC algorithm is the Hybrid Monte-Carlo (HMC) algorithm, which is routinely used for lattice QCD simulations \cite{DeGrandDeTarLQCD,MontvayMuenster}. Recently HMC has also proven to be efficient for QMC simulations of condensed matter systems with non-local inter-electron interactions \cite{Buividovich:13:5,Smekal:13:1,Buividovich:16:1,Assaad:1708.03661}, for which the standard Blankenbecler-Scalapino-Sugar (BSS) QMC scheme \cite{Blankenbecler:PhysRevD.24.2278} requires a large number of auxiliary fields which grows with the number of non-zero potentials at different distances. Furthermore, sometimes local updates of auxiliary fields in the BSS QMC scheme become inefficient in the vicinity of phase transitions \cite{Assaad:1708.03661}. 

The HMC algorithm is most efficient for models with two ``flavours'' of fermionic fields, such as lattice QCD with $N_f = 2$ flavours of light quarks, or tight-binding models with spin-$1/2$ fermions on bipartite lattices at half filling. In such cases the fermionic determinant appears only in the combination $\det\lr{M} \, \det\lr{M^{\dag}} = \det\lr{M M^{\dag}}$, which can be represented in terms of the Gaussian integral over auxiliary ``pseudo-fermion'' fields $Y$:
\begin{equation}
\label{eq:stoch_det}
\det M^\dag M = \int d\bar{Y} dY \, e^{-\bar{Y} (M^\dag M)^{-1} Y} .
\end{equation}
This representation of the fermionic determinant within the HMC algorithm requires the multiplication of the source vector $Y$ by the inverse of the fermionic operator $M^{-1}$, which is equivalent to solving the linear system of equations
\begin{equation}
\label{eq:lin_sys}
 M X = Y .
\end{equation}
In practice, the solution of this linear system takes the largest fraction of CPU time (up to $99 \%$) in HMC simulations.

Yet another situation in which one needs to solve system (\ref{eq:lin_sys}) a large number of times is the stochastic estimation of fermionic observables. For example, the calculation of the trace of a fermionic propagator in determinantal QMC includes a Gaussian stochastic estimator which is the average of the inverse fermionic operator over the set of Gaussian random vectors $Y$:
\begin{equation}
\label{eq:stoch_obser}
 \tr\lr{M^{-1}} = \langle \bar{Y} \, M^{-1} \, Y \rangle_Y.
\end{equation}

Thus the development of efficient solvers for the linear system (\ref{eq:lin_sys}) is an important task relevant for different fields of computational physics. At present HMC codes mostly use iterative solvers, such as various versions of preconditioned Conjugate Gradient, GMRes and BiCGStab algorithms. While iterative solvers are very efficient for well-conditioned sparse matrices, in the vicinity of phase transitions the fermionic matrix $M$ in (\ref{eq:lin_sys}) typically tends to become badly conditioned, which results in the significant growth of the number of iterations and slows down the simulations.

Here we propose an efficient non-iterative solver for the system (\ref{eq:lin_sys}) which is based on the Schur complement method \cite{Schur1917,haynsworth68,HAYNSWORTH196873}. In essence, this solver is a tailored implementation of LU decomposition which takes into account the special band structure of the fermionic matrix, which is typical for interacting tight-binding models in condensed matter physics \cite{Blankenbecler:PhysRevD.24.2278} as well as for staggered \cite{Hasenfratz:NuclPhysB.317.539} and Wilson-Dirac operators used in lattice QCD \cite{Nakamura:1009.2149}.  Despite the fact that the number of floating point operations for our solver scales as the cube of the number of lattice sites,  we find that in the regime of strong correlations (e.g. in the vicinity of a phase transition) it still performs substantially faster than iterative solvers even for rather large lattices. The reason is that the number of operations for our non-iterative solver does not depend on the condition number of the matrix, at least if the numerical round-off errors do not cause any problems in the LU decomposition in the final stage of the algorithm (see below). Thus the method becomes advantageous when the number of iterations in the preconditioned Conjugate Gradient method is significantly increased.

Furthermore, being based on the LU decomposition, this solver benefits from the possibility of re-using the results of all matrix-matrix operations, which can lead to strong performance gains in comparison to the CG method if we need to solve the system (\ref{eq:lin_sys}) for a large number of right-hand side vectors $Y$ and a fixed matrix $M$, for example, in calculations which involve stochastic estimators as in (\ref{eq:stoch_obser}). In this case the number of floating-point operations scales only linearly with the number of lattice sites.

We should also stress that the proposed non-iterative solver, fully analogously to the LU decomposition, is exact if the calculations are performed in infinite precision. Therefore the round-off errors are the only source of inaccuracy and the residual is typically much smaller than the one in solutions obtained from iterative solvers. Thus we practically remove one of the sources of errors from our simulations, which can lead, for instance, to more accurate calculation of the fermionic force during HMC updates of the field configurations. 

The paper is organized as follows. In Section~\ref{sec:sec1} we describe the special band structure of the fermionic matrix $M$ for tight-binding models of interacting fermions, which we will use to describe the practical implementation of the Schur solver. In Section~\ref{sec:sec2} we describe the solver and give a rough estimate of how the complexity of the calculations scales with the lattice size. The solver is described as a general algorithm applicable for any model of lattice fermions, which can be rewritten in the form introduced in Section~\ref{sec:sec1}.  In Section~\ref{sec:sec3} we present the results of numerical tests comparing the Schur solver with preconditioned CG for two particular cases of QMC simulations of interacting fermions on the graphene hexagonal lattice, both with local on-site interactions and non-local (Coulomb) interaction potentials of different strength. In the concluding Section~\ref{sec:summary} we discuss further applications of the developed non-iterative solver.

\section{The structure of the fermionic matrix for tight-binding models of interacting fermions}
\label{sec:sec1}

We start with a brief description of the general structure of the fermionic matrix $M$, considering a general form of the tight-binding model of interacting fermions which is typical in many-body physics.

QMC algorithms usually deal with the path integral representation of the partition function
\begin{equation}
\label{eq:part_func}
\mathcal{Z}=\Tr \exp\lr{-\beta \hat{H}},
\end{equation}
and the corresponding thermodynamic averages of observables
\begin{equation}
\label{eq:observable}
\langle \hat{O} \rangle =\frac{1}{\mathcal{Z}} \Tr( \hat{O} e^{-\beta \hat{H}}).
\end{equation}
Here $\hat H$ and $\hat O$ are the Hamiltonian and some observable respectively and $\beta$ is the inverse temperature. We consider a generic fermionic Hamiltonian $\hat{H}$ with interactions between local fermion charges:
\begin{equation}
\label{eq:full_hamiltonian}
\hat H = \hat H_{(2)} + \hat H_{(4)} =\sum_{ij} h_{ij} {\hat c^\dag_{i}} \hat c_{j} + \sum_{ij} U_{ij} {\hat n_{i}} \hat n_{j},
\end{equation}
where indices $i,j=1...N_s$ are general indices which define lattice sites, spin components, orbitals, etc., and $\hat n_i = {\hat c^\dag_{i}} \hat c_{i}$. The first part in (\ref{eq:full_hamiltonian}) contains only bilinear fermionic terms and is defined by the matrix of one-particle Hamiltonian $h_{ij}$. The second part contains four-fermion terms describing inter-fermion interactions. The most prominent example of such a Hamiltonian is the Hubbard model which includes only local on-site interaction of fermions with different spins.

The path integral representation of the partition function (\ref{eq:part_func}) starts from the Suzuki-Trotter decomposition of the partition function
\begin{equation}
\label{eq:trotter}
  \Tr e^{-\beta \hat H} \approx \Tr \left({ e^{-{\Delta \tau} \, \hat H_{(2)}} e^{-{\Delta \tau} \, \hat H_{(4)}}  e^{-{\Delta \tau} \, \hat H_{(2)}} e^{-{\Delta \tau} \, \hat H_{(4)}} ... } \right)
\end{equation}
into a product of $2 N_t$ exponentials, each associated with a small interval of Euclidean time $\tau \in \lrs{0, \beta}$ of size ${\Delta \tau} \, = \beta/N_t$. The next step is the Hubbard-Stratonovich decomposition of the exponents $e^{-{\Delta \tau} \, \hat H_{(4)}}$ which contain the interaction term:
\begin{eqnarray}
\label{eq:continuous_HS_imag}
  e^{-\frac{{\Delta \tau} \,}{2}\sum_{ij} U_{ij} \hat n_i \hat n_j} 
  \cong 
  \int D \phi_x \, 
  e^{- \frac{1}{2{\Delta \tau} \,} \sum_{ij} \phi_i U^{-1}_{ij} \phi_j} \,
  e^{i \sum_i \phi_i \hat n_i}, 
  \\ \label{eq:continuous_HS_real}
  e^{\frac{{\Delta \tau} \,}{2}\sum_{ij} U_{ij} \hat n_i \hat n_j} 
  \cong 
  \int D \phi_x \,
  e^{- \frac{1}{2{\Delta \tau} \,} \sum_{ij} \phi_j U^{-1}_{ij} \phi_j} \,
  e^{ \sum_i \phi_i \hat n_i} ,
\end{eqnarray}
where we have to use the transformation (\ref{eq:continuous_HS_imag}) for repulsive interactions (opposite charges repel) and (\ref{eq:continuous_HS_real}) for attractive interactions. Upon the Hubbard-Stratonovich transformation all the exponents in the Suzuki-Trotter decomposition (\ref{eq:trotter}) contain only the operators which are quadratic in the fermionic fields (fermionic bilinears), at the expense of introducing an additional path integral over the fields $\phi_i$.

Exponents of the fermionic bilinear operators can be further transformed into the matrix exponents of operators on the much smaller one-particle Hilbert space by using, for example, the formalism of the Grassmann coherent states $\ket{\xi}$ with $\hat{c}_i \ket{\xi} = \xi_i \ket{\xi}$, where the anti-commuting Grassmann variables $\bar{\xi}_i$, $\xi_i$ are in one-to-one correspondence with the fermionic creation and annihilation operators $\hat{c}^{\dag}_i$, $\hat{c}_i$ \cite{MontvayMuenster}. We insert the identity decomposition $I = \int d\bar{\xi} d\xi \, e^{-\bar{\xi} \xi} \, \ket{\xi}\bra{\xi}$ between all the exponentials in (\ref{eq:trotter}) and use the relation
\begin{equation}
\label{eq:grass_av}
 \langle \xi | e^{- {\Delta \tau} \, \sum_{ij} h_{ij} \hat c^{\dag}_i \hat c_j} | \xi' \rangle = e^{\left(   { \sum_{ij} \xi_i (e^{-{\Delta \tau} \, h})_{ij} \xi'_j   }   \right) } .
\end{equation}
After integrating out the Grassmann variables
\begin{equation}
\label{eq:grass_det}
 \int D\bar \xi \xi e^{\left(   { \sum_{ij} \bar \xi_i M_{ij} \xi_j   }   \right) }  = \det M
\end{equation}
we arrive at the following form of the fermionic matrix $M$:
\begin{equation}
\label{eq:M}
 M =
 \begin{pmatrix}
 I           & D_1 &     &        &        &                \\
             & I   & D_2 &        &        &                \\
             &     &  I  & D_3    &        &                \\
             &     &     & \ddots & \ddots &                \\
             &     &     &        & I      &  D_{2N_t-1}    \\
 D_{2N_t}    &     &     &        &        &    I           \\
\end{pmatrix} ,
\end{equation}
where the blocks $D_i$ are $N_s \times N_s$ matrices, similarly to the one-particle Hamiltonian $h_{ij}$. As follows from the Suzuki-Trotter decomposition (\ref{eq:trotter}), the matrix $M$ contains $2 N_t \times 2 N_t$ such blocks. Even blocks are diagonal matrices which contain exponents with auxiliary Hubbard-Stratonovich fields
\begin{equation}
\label{eq:D2k}
 D_{2 k} = \pm
 \begin{pmatrix}
 e^{i \phi^k_1}  &        &                     \\
                 & \ddots &                     \\
                 &        &  e^{i \phi^k_{N_s}} \\
 \end{pmatrix} ,
\end{equation}
where we take plus sign for the last block with $k=N_t$ and minus sign otherwise.

For odd blocks which correspond to the exponents of the fermionic bilinear term in the Hamiltonian (\ref{eq:full_hamiltonian}) one can use two different forms. First, one can use the matrix exponent of a one-particle hamiltonian $h$, as suggested by (\ref{eq:grass_av}):
\begin{equation}
\label{eq:D2k_1_exp}
 D_{2k-1} = -e^{-{\Delta \tau} \, \, h} .
\end{equation}
This form can be advantageous for preserving some symmetries of the original Hamiltonian (\ref{eq:full_hamiltonian}) at the level of the discretized path integral \cite{Buividovich:16:4}. However, since the Trotter decomposition anyway introduces a discretization error of order $O\lr{{\Delta \tau}^2 \,}$ in the partition function (\ref{eq:part_func}) and observables (\ref{eq:observable}), one can also expand the exponential in (\ref{eq:D2k_1_exp}) to the leading order in ${\Delta \tau} \,$:
\begin{equation}
\label{eq:D2k_1_sparse}
 D_{2k-1} = - 1 + {\Delta \tau} \, \, h .
\end{equation}
The advantage of this form is that the blocks $D_{2k-1}$ are sparse matrices, which allows to significantly speed-up matrix-matrix and matrix-vector operations. In practice, we have found that also many elements of the non-sparse matrix (\ref{eq:D2k_1_exp}) are numerically very small, of order $10^{-5}$ and smaller, and can be set to zero without introducing any noticeable error in the results of Monte-Carlo simulations. This allows to use sparse linear algebra to speed up the algorithm even for the exponential representation (\ref{eq:D2k_1_exp}).

However, the algorithm which we describe in this work does not depend on the particular form of the blocks $D_i$, thus in what follows we will work with a general form of the blocks $D_i$ which is valid both for (\ref{eq:D2k_1_exp}) and (\ref{eq:D2k_1_sparse}).

Here we took as an example the Gaussian Hubbard-Stratonovich transformation with only one Hubbard field. In principle, more general decompositions of the interaction term $H_{(4)}$ are also useful \cite{Ulybyshev:1712.02188}, but they do not lead to significant changes in the structure of the matrix (\ref{eq:M}) and all further derivations still remain valid. A detailed discussion of the structure of the fermionic matrix (\ref{eq:M}) can be found in \cite{Buividovich:13:5,Smekal:13:1} for the Hubbard-Coulomb model on the hexagonal lattice, and in \cite{Buividovich:16:1,Assaad:1708.03661} for more general cases.

The band form of the fermionic matrix (\ref{eq:M}) is not unique to the interacting tight-binding models in condensed matter physics. Fermionic matrices (Dirac operators) which are commonly used for lattice QCD simulations can be also represented in the band form (\ref{eq:M}), which is especially useful for lattice QCD simulations based on the canonical partition functions at fixed particle number \cite{Alexandru:1009.2197,Nakamura:1504.04471,Bornyakov:1611.04229}. The transformation to the band form (\ref{eq:M}) is particularly simple for staggered fermions \cite{Hasenfratz:NuclPhysB.317.539}, and requires additional matrix transformations on the blocks $D_i$ for Wilson-Dirac fermions \cite{Nakamura:1009.2149} due to nontrivial dependence on time-like link variables. 

\section{Description of the Schur complement solver}
\label{sec:sec2}

The main idea of the Schur complement solver is the iterative contraction of the number of Euclidean time steps. At each iteration we effectively decrease $N_t$ by a factor of two. In the end we arrive at the matrix of size $\lr{N_s \, \lceil N_t/2^{l_\text{max}}\rceil} \times \lr{N_s \, \lceil N_t/2^{l_\text{max}}\rceil}$, where $l_\text{max}$ is the number of iterations and $\lceil x \rceil$ is the ceiling function. For this much smaller matrix we can already use LU decomposition of sparse matrices in order to solve the remaining linear system.

\begin{algorithm}[h!t]
\caption{Iterative Schur complement solver for arbitrary $N_t$}
\label{alg:schur_solver}
\begin{center}
\begin{algorithmic}[1]
      \Function{$X^{(l)}=$SchurComplementSolve}{$Y^{(l)}$, $N_t^{(l)}$, $D^{(l)}_k$}
      \State // Computes the solution to $ M^{(l)} X^{(l)} = Y^{(l)}$,
      \State // where $M^{(l)}$ has the form (\ref{eq:M}) with off-diagonal blocks $D^{(l)}_k$

      \State $N_t^{(l+1)} := \lceil N_t^{(l)}/2 \rceil $
      \If{$\mod(N_t^{(l)},2)=0$}
      \State $D^{(l+1)}_k := - D^{(l)}_{2 k} D^{(l)}_{2 k + 1}, \quad k = 1 \ldots N_t^{(l+1)}-1$
      \State $D^{(l+1)}_{N_t^{(l+1)}} := - D^{(l)}_{N_t^{(l)}} D^{(l)}_1$
      \State $Y^{(l+1)}_k := Y^{(l)}_{2 k} - D^{(l)}_{2 k} Y^{(l)}_{2 k + 1}, \quad k = 1 \ldots N_t^{(l+1)}-1$
      \State $Y^{(l+1)}_{N_t^{(l+1)}} := Y^{(l)}_{N_t^{(l)}} - D^{(l)}_{N_t^{(l)}} Y^{(l)}_1$
      \ElsIf{$\mod(N_t^{(l)},2)=1$}
      \State $D^{(l+1)}_k := - D^{(l)}_{2 k-1} D^{(l)}_{2 k}, \quad k = 1 \ldots N_t^{(l+1)}-1$
      \State $D^{(l+1)}_{N_t^{(l+1)}} := D^{(l)}_{N_t^{(l)}}$
      \State $Y^{(l+1)}_k := Y^{(l)}_{2 k-1} - D^{(l)}_{2 k-1} Y^{(l)}_{2 k }, \quad k = 1 \ldots N_t^{(l+1)}-1$
      \State $Y^{(l+1)}_{N_t^{(l+1)}} := Y^{(l)}_{N_t^{(l)}}$
      \EndIf
      \If{$l<l_\text{max}$}
      \State $X^{(l+1)}=$\Call{SchurComplementSolve}{$Y^{(l+1)}$, $N_t^{(l+1)}$, $D^{(l+1)}_k$}
      \ElsIf{$l = l_\text{max}$}
      \State Solve $M^{(l+1)} {X}^{(l+1)} = Y^{(l+1)}$ using LU decomposition
      \EndIf
      \If{$\mod(N_t^{(l)},2)=0$}
      \State $X^{(l)}_{2k-1} := Y^{(l)}_{2k-1} - D^{(l)}_{2 k - 1} X^{(l+1)}_k, \quad k = 1 \ldots N_t^{(l+1)}$
      \State $X^{(l)}_{2 k} := X^{(l+1)}_k, \quad k = 1 \ldots N_t^{(l+1)}$
      \ElsIf{$\mod(N_t^{(l)},2)=1$}
      \State $X^{(l)}_{2k} := Y^{(l)}_{2k} - D^{(l)}_{2 k } X^{(l+1)}_{k+1}, \quad k = 1 \ldots N_t^{(l+1)}-1$
      \State $X^{(l)}_{2 k-1} := X^{(l+1)}_k, \quad k = 1 \ldots N_t^{(l+1)}$
      \EndIf
      \State \Return $X^{(l)}$
      \EndFunction
    \end{algorithmic}
  \end{center}
\end{algorithm}

Similarly to the matrix $M$ we divide the vectors $X$ and $Y$ in (\ref{eq:lin_sys}) into the blocks  of the size $N_s$.
\begin{equation}
  \label{eq:v_blocks}
  X \equiv X^{(1)}=
  \begin{pmatrix}
    X^{(1)}_1 \\ X^{(1)}_2 \\ \vdots  \\ X^{(1)}_{K_1-1} \\ X^{(1)}_{K_1}
  \end{pmatrix}
  , \qquad
  Y \equiv Y^{(1)}=
  \begin{pmatrix}
    Y^{(1)}_1 \\ Y^{(1)}_2 \\ \vdots  \\ Y^{(1)}_{K_1-1} \\ Y^{(1)}_{K_1}
  \end{pmatrix}
  .
\end{equation}
In here, the upper index denotes the Schur iteration. At $l$\textsuperscript{th} iteration the number of blocks $K_l$ decreases as $K_{l+1} = \lceil K_l/2 \rceil$, where $K_1 = 2 N_t$ corresponds to the original system (\ref{eq:lin_sys}), which is now written as
\begin{equation}
  \label{eq:lin_sys1}
  M^{(1)} X^{(1)} = Y^{(1)},
\end{equation}
where $M^{(1)}$ is the equivalent of the initial matrix (\ref{eq:M}) in which the blocks $D^{(1)}_k$ also acquire the upper index according to the number of Schur iteration being performed.

We start with the iteration $l = 1$, for which $$\mod\lr{K_l,2} = \mod\lr{2 N_t, 2} = 0$$ and the matrix $M$ in (\ref{eq:lin_sys}) has the form (\ref{eq:M}). At the beginning of each iteration, we first perform a permutation of blocks of size $N_s$ inside the vector $X$:
\begin{equation}
  \label{eq:perm}
  P_{K_l}
  \begin{pmatrix}
    X_1 \\ X_2 \\ \vdots  \\ X_{K_{l-1}} \\ X_{K_l}
  \end{pmatrix}
  =
  \begin{pmatrix}
    X_1 \\ X_{K_l/2+1} \\ X_2\\ X_{K_l/2+2} \\ \vdots  \\ X_{K_l/2} \\ X_{K_l}
  \end{pmatrix}
  .
\end{equation}

We also apply this permutation both to all entities (matrix, vectors) in (\ref{eq:lin_sys1}):
\begin{equation}
  \label{eq:lin_sys_perm}
  P_{K_l}^\dag M^{(l)} P_{K_l} \, P_{K_l}^\dag X^{(l)} = P_{K_l}^\dag Y^{(l)}.
\end{equation}
The inverse permutation $P_{K_l}^{\dag}$ looks like:
\begin{equation}
  \label{eq:perm_inv}
  P_{K_l}^{\dag}
  \begin{pmatrix}
    X_1 \\ X_2 \\ \vdots  \\ X_{K-1} \\ X_{K_l}
  \end{pmatrix}
  =
  \begin{pmatrix}
    X_1 \\ X_3 \\ \vdots \\ X_{K_l-1} \\ X_2 \\ X_4\\ \vdots  \\ X_{K_l}
  \end{pmatrix}
  .
\end{equation}
After the permutation of columns and rows is applied to the matrix $M^{(l)}$, it takes the following form:
\begin{eqnarray}
  \label{eq:matrix_bar}
  P_{K_l}^\dag M^{(l)} P_{K_l} \equiv \bar{M}^{(l)}
  = \nonumber \\ =
  \begin{pmatrix}
    I             &           &           &        & D^{(l)}_1 &           &        &                 \\
                  & I         &           &        &           & D^{(l)}_3 &        &                 \\
                  &           & \ddots    &        &           &           & \ddots &                 \\
                  &           &           & I      &           &           &        & D^{(l)}_{K_l-1} \\
    0             & D^{(l)}_2 &           &        & I         &           &        &                 \\
                  & \ddots    & D^{(l)}_4 &        &           & I         &        &                 \\
                  &           & \ddots    & \ddots &           &           & \ddots &                 \\
    D^{(l)}_{K_l} &           &           & 0      &           &           &        & I               \\
  \end{pmatrix}
  .
\end{eqnarray}
Thus, $\bar{M}^{(l)}$ can be then divided into four large blocks of the same size:
\begin{equation}
  \label{eq:matrix_bar_blocks}
  \bar{M}^{(l)}
  =
  \begin{pmatrix}
    I       & R^{(l)} \\
    Q^{(l)} & J^{(l)} \\
  \end{pmatrix}
  ,
\end{equation}
where $J^{(l)}=I$,
\begin{equation}
  \label{eq:R_block}
  R^{(l)} =
  \begin{pmatrix}
    D^{(l)}_1 &           &        &                 \\
              & D^{(l)}_3 &        &                 \\
              &           & \ddots &                 \\
              &           &        & D^{(l)}_{K_l-1} \\
  \end{pmatrix}
  ,
\end{equation}
and
\begin{equation}
  \label{eq:Q_block}
  Q^{(l)} =
  \begin{pmatrix}
    0 & D^{(l)}_2 &        &   &   \\
      & \ddots    & \ddots &   &   \\
    &               &   \ddots      &  D^{(l)}_{K_l-2}     \\
    D^{(l)}_{K_l}      &       &        &  0  \\
  \end{pmatrix}
  .
\end{equation}
After similar division of permuted vectors into upper and lower components
\begin{equation}
  \label{eq:xy_bar}
  {\bar X}^{(l)} \equiv P_{K_l}^\dag X^{(l)} =
  \begin{pmatrix}{U_X}^{(l)} \\ {L_X}^{(l)}
  \end{pmatrix}
  , \quad{\bar Y}^{(l)} \equiv P_{K_l}^\dag Y^{(l)} =
  \begin{pmatrix}{U_Y}^{(l)} \\ {L_Y}^{(l)}
  \end{pmatrix}
  ,
\end{equation}
where each component contains $K_l/2$ blocks of size $N_s$, the system (\ref{eq:lin_sys_perm}) takes the form
\begin{equation}
  \label{eq:lin_sys_new1}
  \begin{cases}
    \phantom{Q^{(l)}} U_X^{(l)} + R^{(l)} L_X^{(l)}  =  U_Y^{(l)} , &   \\
    Q^{(l)} U_X^{(l)} + J^{(l)} L_X^{(l)}           =  L_Y^{(l)} .  &   \\
  \end{cases}
\end{equation}
Expressing $U_X^{(l)}$ from the first equation in (\ref{eq:lin_sys_new1}) as
\begin{equation}
  \label{eq:Ux_vs_Uy}
  U_X^{(l)} = U_Y^{(l)} - R^{(l)} L_X^{(l)}
\end{equation}
we can rewrite the second equation in (\ref{eq:lin_sys_new1}) as
\begin{equation}
  \label{eq:lin_sys_new}
  \lr{J^{(l)} - Q^{(l)}R^{(l)}} L_X^{(l)} = L_Y^{(l)}- Q^{(l)} U_Y^{(l)}  .
\end{equation}
Once we solve equation (\ref{eq:lin_sys_new}) and find ${L_X}^{(l)}$, the upper part $U_X^{(l)}$ of the vector ${\bar X}^{(l)}$ can be immediately found using the first equation in (\ref{eq:lin_sys_new1}). Thus we have effectively reduced the size of the linear system (\ref{eq:lin_sys_perm}) by a factor of two.

An important observation is that the Schur complement of $\bar{M}^{(l)}$, i.e., the matrix $\lr{J^{(l)} - Q^{(l)}R^{(l)}}$, which appears in the equation (\ref{eq:lin_sys_new}) has exactly the same form as the original matrix $M \equiv M^{(1)}$ in (\ref{eq:M}):
\begin{eqnarray}
  \label{eq:M_next_iter}
  \lr{J^{(l)} - Q^{(l)}R^{(l)}}
  =
  \begin{pmatrix}
    I                   & D^{(l+1)}_1 &             &        &                       \\
                        & I           & D^{(l+1)}_2 &        &                       \\
                        &             & \ddots      & \ddots &                       \\
                        &             &             & I      & D^{(l+1)}_{K_{l+1}-1} \\
    D^{(l+1)}_{K_{l+1}} &             &             &        & I                     \\
  \end{pmatrix}
\end{eqnarray}
with
\begin{eqnarray}
  \label{eq:D_iter}
  D^{(l+1)}_k & := & - D^{(l)}_{2k}  D^{(l)}_{2k+1}, \quad k = 1 \ldots K_{l+1}-1, \nonumber \\
  D^{(l+1)}_{K_{l+1}} & := & - D^{(l)}_{K_l} D^{(l)}_{1}.
\end{eqnarray}

Now we can repeat the same steps (\ref{eq:lin_sys_perm}) - (\ref{eq:lin_sys_new}) with the following substitution:
\begin{eqnarray}
  \label{eq:full_iter_step}
  M^{(l+1)} & := & J^{(l)} -Q^{(l)}R^{(l)}, \nonumber \\
  X^{(l+1)} & := & {L_X}^{(l)}, \nonumber \\
  Y^{(l+1)} & := & {L_Y}^{(l)}- Q^{(l)} {U_Y}^{(l)} .
\end{eqnarray}

The case $\mod(K_l,2)=1$ is treated as follows. We first artificially increase the size of the system $M^{(l)}X^{(l)} = Y^{(l)}$ by the blocksize $N_s$ and thus consider
\begin{eqnarray}
  \label{eq:linear_system_prime}
  {M}'^{(l)} {X}'^{(l)} = {Y}'^{(l)}
\end{eqnarray}
which is now a system of blocksize ${K}_l' \times {K}_l',$ $ {K}_l' = K_l+1$ with
\begin{eqnarray}{M}'^{(l)} =
  \begin{pmatrix}
    I & 0       \\
    0 & M^{(l)}
  \end{pmatrix}
  \text{ and }
  {Y}'^{(l)} =
  \begin{pmatrix}
    0 \\ Y^{(l)}
  \end{pmatrix}
  .
\end{eqnarray}
We then proceed by applying the steps (\ref{eq:lin_sys_perm}) - (\ref{eq:lin_sys_new}) to the modified system (\ref{eq:linear_system_prime}). The permutation of ${M}'^{(l)}$ leads to
\begin{equation}
  \label{eq:matrix_bar_prime}
  P_{{K}_l'}^\dag {{M}'^{(l)}} P_{{K}_l'} \equiv \bar{M}'^{(l)}
  =
  \begin{pmatrix}
    I        & R'^{(l)} \\
    Q'^{(l)} & J'^{(l)} \\
  \end{pmatrix}
  ,
\end{equation}
where again $I$, $J'^{(l)}$, $R'^{(l)}$ and $Q'^{(l)}$ are of the same blocksize $K_l'/2 \times K_l'/2$ with
\begin{equation}
  \label{eq:J_and_R_block_prime}
  J'^{(l)}=
  \begin{pmatrix}
    I \\
    & \ddots \\
    D^{(l)}_{K_l} &   & I \\
  \end{pmatrix}
  ,\quad
  R'^{(l)} =
  \begin{pmatrix}
    0 &           &        &                 \\
    & D^{(l)}_2 &        &                 \\
    &           & D^{(l)}_4 &                 \\
    &           &        & \ddots \\
      &   &   &   & D^{(l)}_{K_l-1} \\
  \end{pmatrix}
  ,
\end{equation}
and
\begin{equation}
  \label{eq:Q_block_prime}
  Q'^{(l)} =
  \begin{pmatrix}
    0 & D^{(l)}_1 &        &   &   \\
      & \ddots    & \ddots &   &   \\
    &        & \ddots & D^{(l)}_{K_l-2} \\
    &        &        & 0               \\
  \end{pmatrix}
  .
\end{equation}
The Schur complement $\lr{J'^{(l)} - Q'^{(l)}R'^{(l)}}$ has again the same structure as in (\ref{eq:M_next_iter}) with
\begin{eqnarray}
  \label{eq:D_iter_prime}
  D^{(l+1)}_k & := & - D^{(l)}_{2k-1}  D^{(l)}_{2k}, \quad k = 1 \ldots K_{l+1}-1, \nonumber \\
  D^{(l+1)}_{K_{l+1}} & := &  \phantom{-}D^{(l)}_{K_l}.
\end{eqnarray}
Thus we can proceed with iterations switching between eq. (\ref{eq:D_iter}) and (\ref{eq:D_iter_prime}) if necessary. 

If the number of Euclidean time slices $N_t$ is some power of two, i.e., $N_t=2^m$, the matrix $M^{(m+1)}$ takes the form
\begin{equation}
  \label{eq:M_final}
  M^{(m+1)} =
  \begin{pmatrix}
    I                                        & (-1)^m \prod^{N_t}_{k=1} D^{(1)}_k \\
    (-1)^m \prod^{N_t}_{k=1} D^{(1)}_{k+N_t} & I                                  \\
  \end{pmatrix}
  .
\end{equation}
The final linear system (the second equation in the system (\ref{eq:lin_sys_new})) which we need to solve involves the matrix:
\begin{equation}
  \label{eq:IQR_final}
  I - Q^{(m+1)} \, R^{(m+1)}
  =
  I - \prod^{2N_t}_{k=1} D^{(1)}_k.
\end{equation}
In contrast to the original system of linear equations (\ref{eq:lin_sys}) with $\lr{N_s N_t} \times \lr{N_s N_t}$ matrix $M$, the system (\ref{eq:IQR_final}) involves only much smaller $N_s \times N_s$ matrices and thus can be solved using non-iterative (exact) methods such as the LU decomposition.

The solution $X^{(l_\text{max})}$ is finally obtained in the last iteration, after the LU decomposition of the $M^{(l_\text{max})}$ matrix. Subsequently, we can revert all iterations using the following relations:
\begin{equation}
  \label{eq:X_iter_back}
  X^{(l-1)} = P_{l-1}
  \begin{pmatrix}{U_Y}^{(l-1)}- R^{(l-1)}  X^{(l)} \\ X^{(l)}
  \end{pmatrix}
  .
\end{equation} 
In the case of odd blocksize $\mod(K_{l-1},2)=1$, this step yields ${X'}^{(l-1)}$ because of the artificial enlargement of the linear system described by equation (\ref{eq:linear_system_prime}). Thus in order to obtain $X^{(l-1)}$ we have to omit the first block of size $N_s$ from ${X'}^{(l-1)}$. In the end we restore the initial vector $X \equiv X^{(1)}$. 

Algorithm~\ref{alg:schur_solver} summarizes the above description of the iterative Schur complement solver in terms of pseudocode, where all permutations $P_l$ and $P_l^{\dag}$ are made explicit. 

In practice the condition number of the matrices $D^{(l)}_k$ grows exponentially with $l$. As a result, we found a hard limit of $l_\text{max}=6-8$ Schur iterations using double precision due to the fact that the matrices $D^{(l)}_k$ with $l>l_\text{max}$ cannot be calculated and expressed in this floating-point format. $l_\text{max}$ can be increased  by a small number of iterative refinement steps \cite{Moler:ACM1967} as described in the Algorithm~\ref{alg:IterativeRefinement}. For $l\leq l_\text{max}$, the iterative refinement can be used for suppression of round-off error in the bare Schur complement solver.

\begin{algorithm}[h!t]
\caption{Iterative Refinement}
\label{alg:IterativeRefinement}
\begin{center}
\begin{algorithmic}[1]
      \Function{$X=$IterativeRefinement}{$M,Y$}
      \State // Computes the solution to $ M X = Y$
      \State $X:=0$, $R:=Y$
      \While{Solution $X$ too inaccurate}
      \State Solve $MU=R$ via Schur complement solver
      \State Add the correction to the solution $X:=X+U$ 
      \State Calculate new residual $R:=Y-MX$
      \EndWhile
     \EndFunction
\end{algorithmic}
\end{center}
\end{algorithm}

Note that, if we work with initially sparse above-diagonal blocks $D^{(1)}_k$ of the form (\ref{eq:D2k_1_sparse}), the blocks $D^{(l)}_k$ become less and less sparse as the number of Schur iterations $l$ increases. As a result, sparse matrix multiplications become less and less efficient.

Clearly, one needs to store the matrix blocks $D^{(l)}_k$ in memory in order to perform backward iterations which reconstruct $X^{(l-1)}$ from $X^{(l)}$. Thus the method is rather memory consuming. On the other hand, once all matrix blocks $D^{(l)}_k$ are calculated and the final LU decomposition of the final matrix $M^{(l_\text{max})}$ is performed, all further solutions of the linear system (\ref{eq:lin_sys}) with different right-hand sides $Y$ become very cheap, since only a limited number of matrix-vector operations is needed in order to perform full reconstruction of the solution $X$ using equation (\ref{eq:X_iter_back}). The complexity of these matrix-vector operations is comparable to the single application of the original fermionic matrix $M$ to a vector. In particular, this allows to efficiently implement the iterative refinement algorithm \cite{Moler:ACM1967} (see Algorithm~\ref{alg:IterativeRefinement}), since matrix-matrix operations are performed only during the first iteration. Furthermore, one can use the Schur complement solver for the fast inversion of the matrix $M$, which amounts to solving system (\ref{eq:lin_sys}) with multiple right-hand-side vectors $Y$ which have only a single unit element. This sparseness of $Y$ allows one to further speed-up the inversion algorithm.

We can also re-use the results of matrix-matrix multiplications performed when solving the system (\ref{eq:lin_sys}) in order to solve the system of the form 
\begin{equation}
\label{eq:lin_sys_dag}
 M^{\dag} X = Y .
\end{equation}
This possibility is very important in practice, because the system  
\begin{equation}
\label{eq:lin_sys_real}
 M^\dag M X=Y
\end{equation}
naturally appears in HMC calculations of spin-$1/2$ systems, and requires consecutive solution of the systems (\ref{eq:lin_sys}) and (\ref{eq:lin_sys_dag}). For the system (\ref{eq:lin_sys_dag}) the forward iterations take the following form:
\begin{eqnarray}
  \label{eq:dag_iter_forward}
  X^{(l+1)} & := & {L_X}^{(l)}, \nonumber \\
  Y^{(l+1)} & := & {L_Y}^{(l)}- (R^{(l)})^{\dag} {U_Y}^{(l)} ,
\end{eqnarray}
where vectors ${U_X}^{(l)}$, ${L_X}^{(l)}$ and ${U_Y}^{(l)}$, ${L_Y}^{(l)}$ are defined according to (\ref{eq:xy_bar}). In the final iteration we obtain the matrix $(M^{(l_\text{max})})^{\dag}$, thus the LU decomposition can be re-used too. Correspondingly, backward iterations are changed as
\begin{equation}
  \label{eq:dag_iter_back}
  X^{(l-1)} = P_{l-1}
  \begin{pmatrix}{U_Y}^{(l-1)}- (Q^{(l-1)})^\dag  X^{(l)} \\ X^{(l)}
  \end{pmatrix}
  .
\end{equation}
The matrices $Q^{(l)}$ and $R^{(l)}$ are taken from the previous run of the algorithm for the linear system (\ref{eq:lin_sys}) with matrix $M$. 

The number of floating-point operations for the Schur complement solver with initially sparse blocks $D^{(1)}_{2k-1}$ of the form (\ref{eq:D2k_1_sparse}) in (\ref{eq:M}) can be estimated as
\begin{equation}
\label{eq:scaling}
 N_{op} 
 = 
 \sum_{l=1}^{l_\text{max}} \, N_l^2 \, N_s \, \frac{N_t}{2^l}
 + 
 N_{LU} ,
\end{equation}
where $l_{\text{max}}$ is the total number of Schur iterations which is limited either by $log_2\lr{N_t}$ or due to the accumulation of numerical round-off errors, as discussed above. We have also assumed that $N_t = 2^m$ with some positive integer $m$. $N_l$ is the number of non-zero elements in each column (row) of the blocks $D^{(l)}_k$ at $l$-th Schur iteration, which typically grows with $l$ as
\begin{equation}
\label{eq:Nl_def}
 N_l =
 \begin{cases}
   A \, {l}^d, & A \, {l}^d < N_s , \\
   N_s,        & A \, {l}^d > N_s , \\
 \end{cases} 
\end{equation}
where $d$ is the number of spatial lattice dimensions and $A$ is some numerical pre-factor which depends on the number of fermion components, number of nearest neighbors on the lattice of a given type, and so on. For dense matrices, $N_l = N_s$ for all $l$. $N_{LU}$ is the number of floating-point operations required for the LU decomposition, which in general scales with $N_s$ and $N_t$ as 
\begin{equation}
\label{eq:NLU_scaling}
 N_{LU} \sim \lr{N_s \, \frac{N_t}{2^{l_\text{max}}}}^3 .
\end{equation}
While this scaling can be made milder by using sparse linear algebra, it still dominates the CPU time for sufficiently large $N_s$ or $N_t$.

For practical simulations we have implemented three different versions of the Schur complement solver:
\begin{itemize}
 \item A CPU version with sparse linear algebra (intended for use with sparse initial blocks $D_{2k-1}$ of the form (\ref{eq:D2k_1_sparse}))
 \item A CPU version with dense linear algebra (intended for use with dense initial blocks $D_{2k-1}$ of the form (\ref{eq:D2k_1_exp}))
 \item A GPU version with dense linear algebra.
\end{itemize} 
The usage of sparse linear algebra is still advantageous even taking into account the fact that initially sparse blocks $D^{l}_k$ become denser and denser after repeating Schur iterations. We can make the following estimate: in the case of sparse initial blocks $D_{2k-1}$ on 2D hexagonal lattice with spatial dimensions $12\times12$, blocks $D^{l}_k$ still have half of their elements equal to zero even after $l=6$ Schur iterations. Thus the sparse matrix-matrix operations give some speedup even during the last iteration. In addition, if $\log_2\lr{N_t} > l_\text{max}$, or if $N_t$ is not a power of two, the final matrix $M^{(l_\text{max})}$ contains several zero $N_s \times N_s$ blocks away from the diagonal, and (obviously sparse) identity matrices on the diagonal.

However, if the lattice is small enough, the blocks $D^{l}_k$ become dense too fast, even if they were sparse initially. In this situation the dense linear algebra is advantageous in any case. 

For the CPU version of our solver, the final LU decomposition of the matrix $M^{(l_\text{max})}$ is made using the \texttt{SuperLU} library \cite{li05}, which offers very efficient implementations of LU decomposition both for dense and sparse matrices. In the GPU version, we currently work only with dense matrices and use the \texttt{cuSOLVER} library for the final LU decomposition and \texttt{cuBLAS} for the linear algebra operations. Efficient GPU implementation of LU decomposition for sparse matrices have became available only recently \cite{GPU_sparse_LU}, thus we were not yet able to test it.

\section{Numerical tests}
\label{sec:sec3}

As a practical benchmark test for our algorithm, we consider the HMC simulations of the interacting tight-binding model (\ref{eq:full_hamiltonian}) on the hexagonal graphene lattice. The one-particle Hamiltonian in (\ref{eq:full_hamiltonian}) is spin-independent, and its matrix elements $h_{x,y}$ are equal to $-\kappa$ if $x$ and $y$ are neighboring lattice sites, and zero otherwise.

We consider two physically different options for the inter-fermion interaction potential in (\ref{eq:full_hamiltonian}). The first is the on-site interaction potential $U_{xy} = U \delta_{xy}$, which corresponds to the Hubbard model on the hexagonal lattice. At the critical value $U_c = 3.8 \kappa$, this model undergoes a quantum phase transition towards an insulating, strongly correlated anti-ferromagnetic state \cite{Assaad:1304.6340,Buividovich:16:4}.

The second option is the realistic inter-electron interaction potential for suspended graphene \cite{Wehling:11:1}, which was used in HMC simulations in \cite{Buividovich:13:5,Smekal:13:3}. We refer to the latter choice as the Hubbard-Coulomb model. If one uniformly scales this potential as $U_{x y} \rightarrow c \, U_{x y}$, this model also undergoes a phase transition towards an insulating state at $c \approx 1.43 \approx \lr{0.7}^{-1}$ \cite{Buividovich:13:5,Smekal:13:3}. To make a meaningful comparison between the Hubbard and the Hubbard-Coulomb models, we thus use the value $U/U_c = 0.7$ to define the weak-coupling regime.

To this end we select ten configurations of the Hubbard fields $\phi^k_x$ generated by a well-thermalized HMC simulation and use them inside the even blocks $D^{2k}$ as defined in (\ref{eq:D2k}) in order to construct the matrix (\ref{eq:M}). To make a meaningful comparison with the CG algorithm, which is efficient only for sparse matrices, we use the originally sparse form (\ref{eq:D2k_1_sparse}) of the matrix blocks $D_{2 k - 1} \equiv D^{(1)}_{2 k - 1}$.

\begin{figure}[h!tb]
  \centering
  \begin{minipage}{.49\textwidth}
  \includegraphics[width=\textwidth]{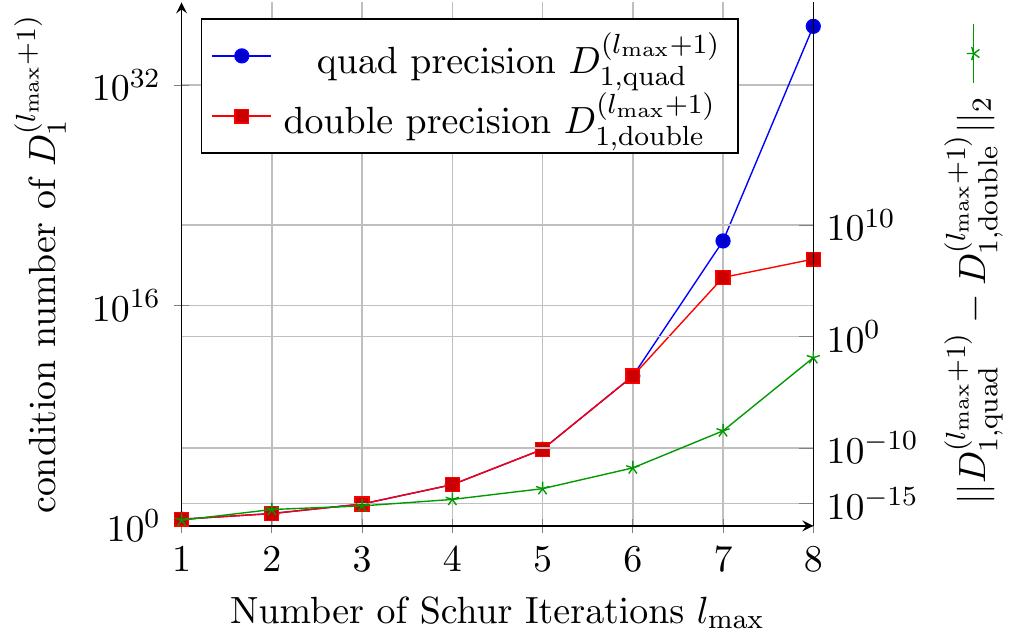}
  \end{minipage}\hfill
    \begin{minipage}{.49\textwidth}
  \includegraphics[width=\textwidth]{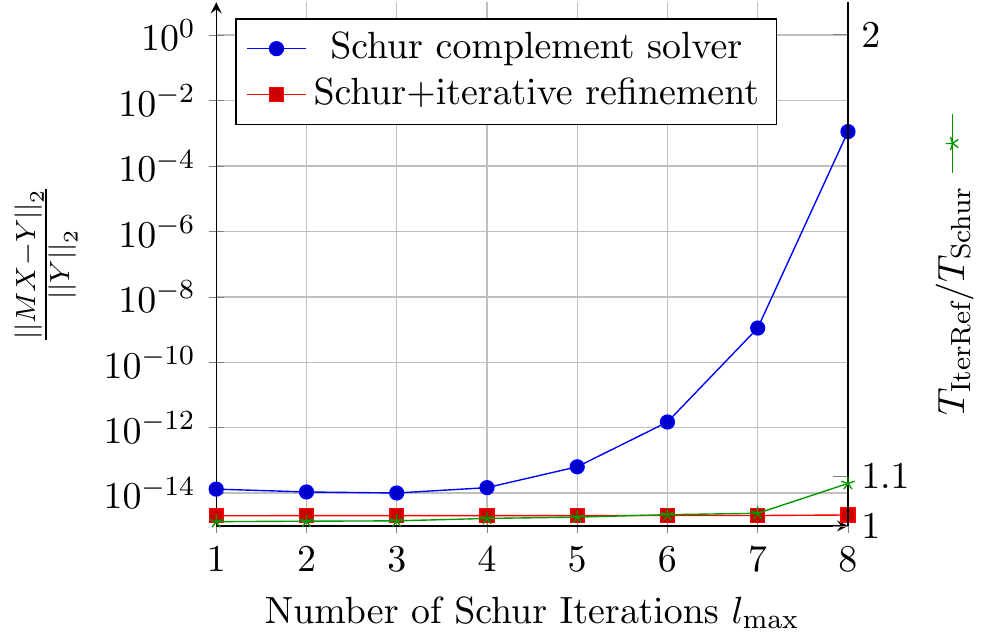}
  \end{minipage}
  \caption{Left: Condition number of the matrices $D_{1}^{l_\text{max}+1}$ in quadruple and double precision as well as the difference of the two variants measured in the two-norm. Right: Comparison of the accuracy and CPU-time of the standalone Schur complement method versus the Schur complement method plus iterative refinement. Here we use one of the test configurations in the strong-coupling phase of the Hubbard model with  $N_s=288$ and $N_t=256$.}
  \label{fig:accuracy_Ns288_Nt128}
\end{figure}

A first set of tests addresses the accuracy of the proposed Schur complement solver. Recall that all matrices $D_{k}^{(l)}$, $l\leq l_{\text{max}}+1$ are used in the Schur complement solver with $l_{\text{max}}$ steps. For the left plot of Fig.~\ref{fig:accuracy_Ns288_Nt128} we first calculated the matrices $D_{1}^{(l_\text{max}+1)}$ in quadruple ($D_{1,\text{quad}}^{(l_\text{max}+1)}$) and double precision ($D_{1,\text{double}}^{(l_\text{max}+1)}$) for one of our test configurations in the strong-coupling phase of the Hubbard model. As already stated in the Section~\ref{sec:sec2}, the condition number of the matrices $D_k^{(l)}$ grows exponentially in $l$ as depicted by plot of the condition number of $D_{1,\text{quad}}^{l_\text{max}+1}$. Due to this fact, the exactness of the matrices $D_k^{(l)}$ calculated in double precision decreases exponentially in comparison to the matrices calculated in quadruple precision which is shown in the plot of $||D_{1,\text{quad}}^{(l_\text{max}+1)}-D_{1,\text{double}}^{(l_\text{max}+1)}||_2$. Therefore, the accuracy of a single Schur complement solution in double precision also decreases exponentially as can be seen in the right plot of Fig.~\ref{fig:accuracy_Ns288_Nt128}. Nevertheless, a relative residual up to machine precision could be obtained via iterative refinement, see Algorithm~\ref{alg:IterativeRefinement}. For $l_\text{max} \leq 7$ the solution was obtained with $2$ steps of iterative refinement whereas $5$ steps were needed in the case $l_\text{max}=8$. For $l_\text{max} > 8$ the method completely failed. The ratio $T_{\text{IterRef}}/T_{\text{Schur}}$ of CPU-time furthermore indicates that the main effort of the Schur method lies in the preparation phase. 

In the next set of tests we compare the timing of the Schur complement solver with that of the Conjugate Gradient iterative algorithm. In both cases (CG and Schur solver) we solve the system (\ref{eq:lin_sys_real}). 
For the Schur solver this means that all matrix-matrix operations and LU decomposition are made once, but matrix-vector operations (which are very cheap) are made twice, first for $M^{\dag}$ and subsequently for $M$ operator, as discussed in the Section~\ref{sec:sec2}. We then calculate the ratio $T_{CG}/T_{Schur}$ of CPU times required to solve the linear system (\ref{eq:lin_sys}) with both methods. While for the Schur complement solver this time depends only on the lattice size, for the Conjugate Gradient method the condition number of the matrix strongly affects the number of iterations required to find the solution with a given precision. 
We use the global relative error 
\begin{equation}
\label{eq:relative_error_def}
 E=\frac{ {||M^{\dag} M X -Y||}_2 }{ {||Y||}_2}
\end{equation}
as a measure of uncertainty during the solution of the system. CG iterations stops when $E<10^{-9}$. This precision is usually enough in HMC calculations. On the other hand, typical uncertainty observed after the application of the Schur solver is much smaller and lies in the region $E=10^{-12}...10^{-11}$. It reflects the ``exact'' nature of the non-iterative solver: the only source of inaccuracy is the round-off errors. 

\begin{figure}[t!]
    \centering
	\includegraphics[width=0.49\textwidth]{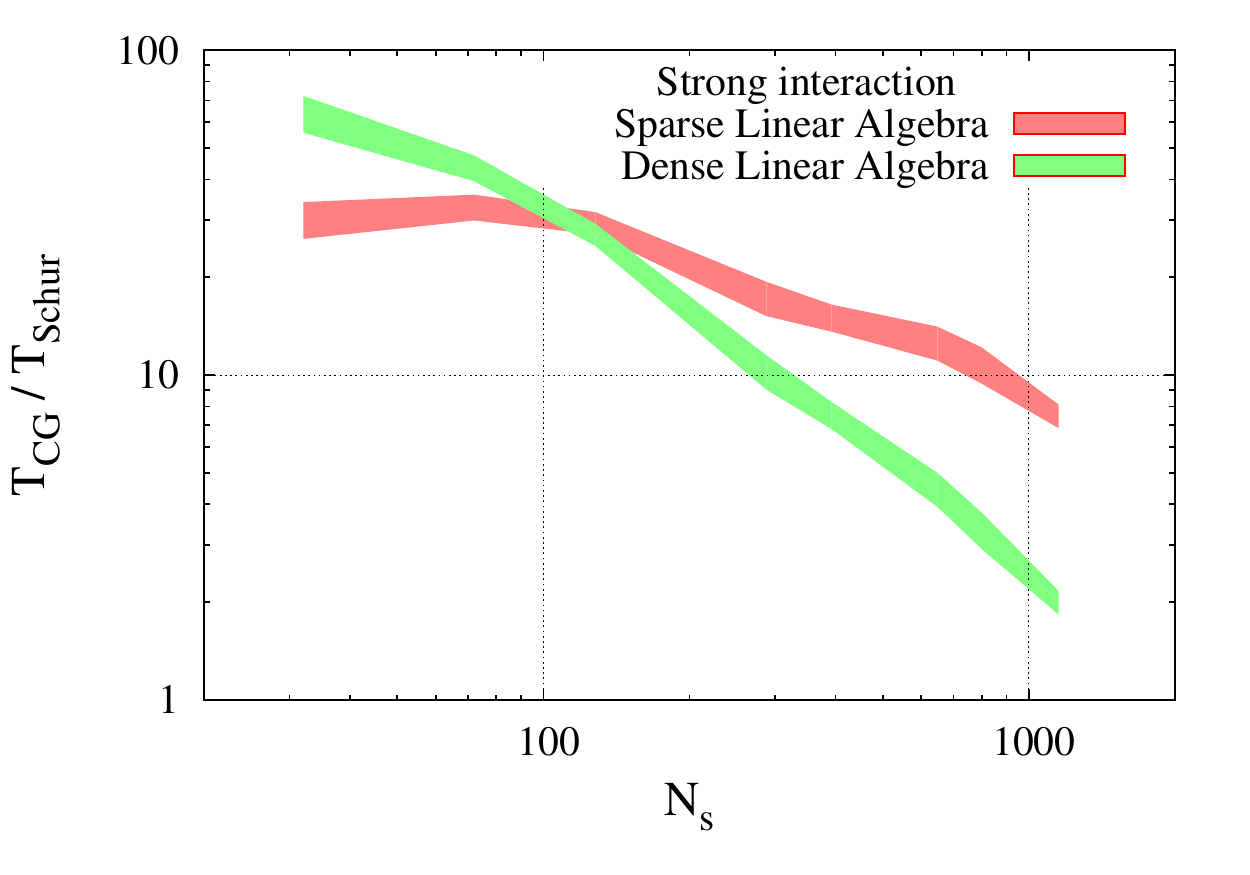}
	\includegraphics[width=0.49\textwidth]{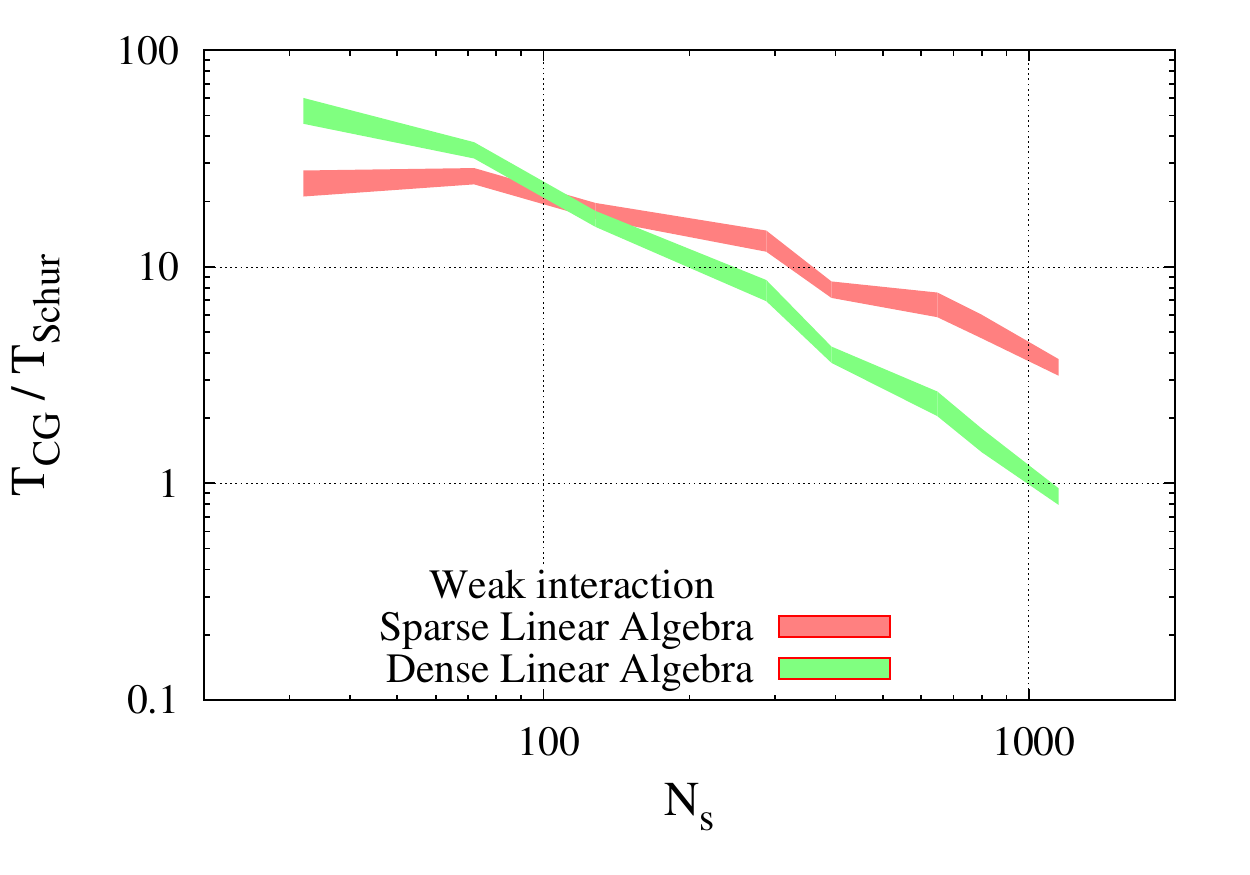}\\
    \includegraphics[width=0.49\textwidth]{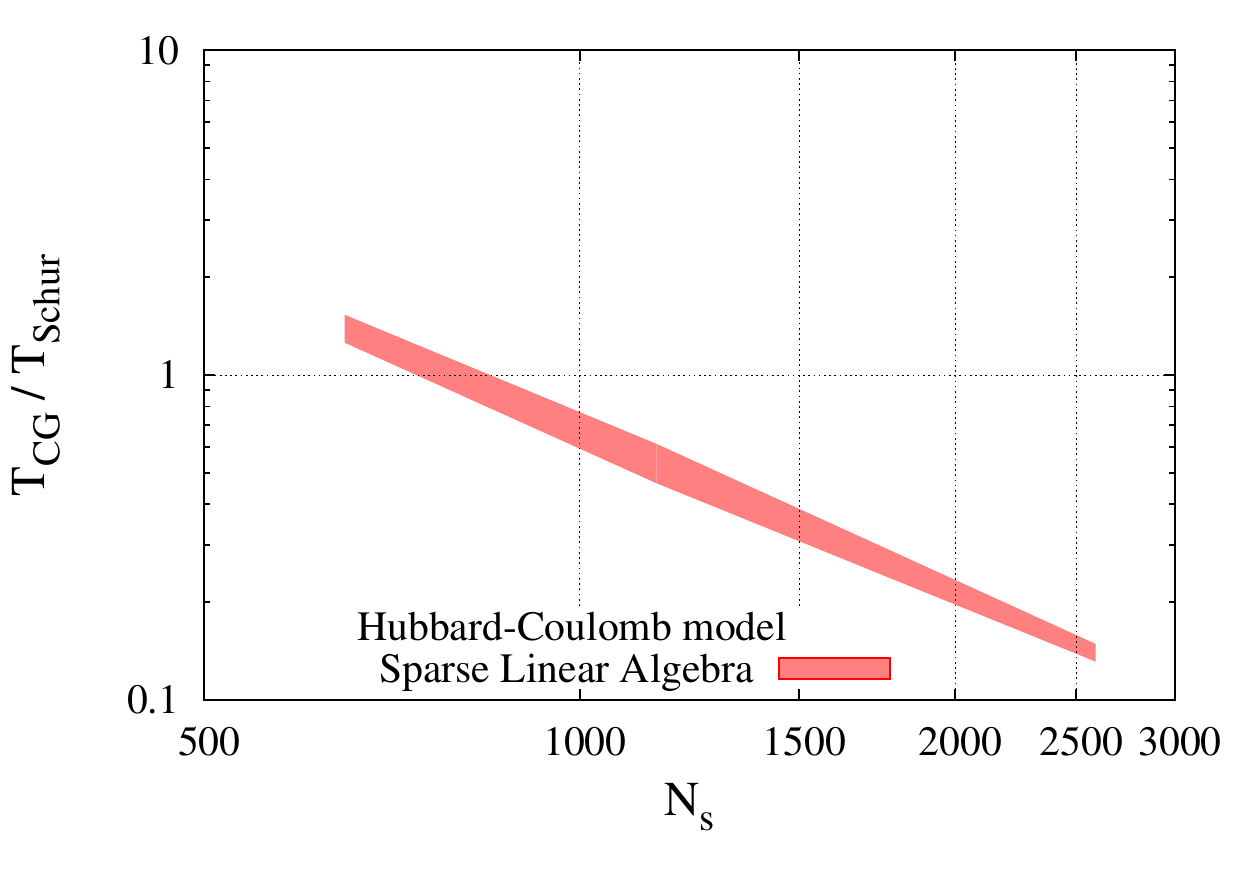}
    \includegraphics[width=0.49\textwidth]{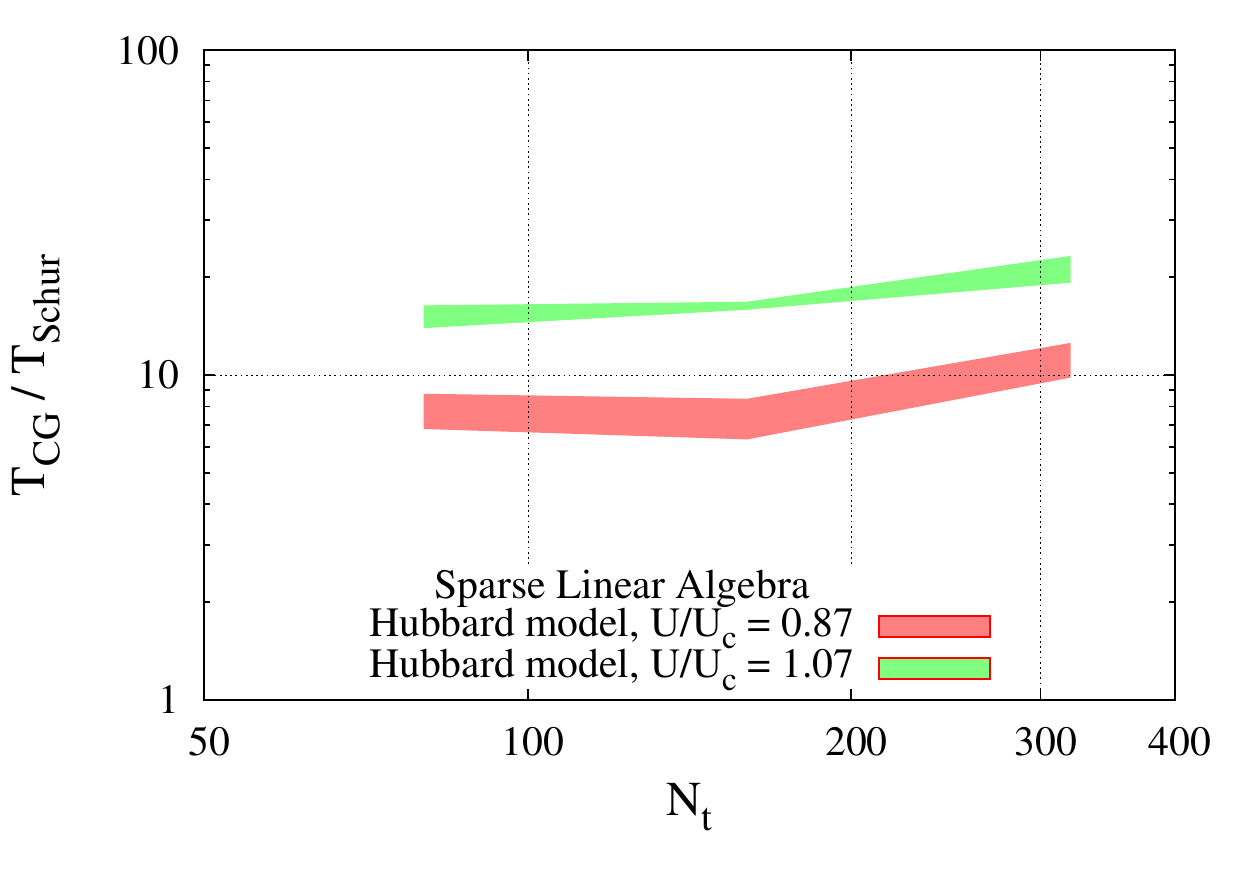}\\
	\caption{Comparison of timings $T_{CG}$ and $T_{Schur}$ of Conjugate Gradient and Schur complement solver for a single system of equations (\ref{eq:lin_sys_real}) for the tight-binding models on the hexagonal lattice. On the top: for the Hubbard model in the strong-coupling regime with $U/U_c = 1.07$ (left plot) and in the weak-coupling regime with $U/U_c = 0.87$ (right plot) at fixed $N_t = 128$ and different $N_s$. Bottom left: for the Hubbard-Coulomb model in the weak coupling regime away from the phase transition ($U/U_c=0.7$) at fixed $N_t = 80$ and different $N_s$. Bottom right: for the Hubbard model at fixed $N_s = 392$ and physical temperature $\beta\kappa=21.6$ and different $N_t$. In most cases we show two curves where either only dense or only sparse linear algebra was used for all matrix-matrix multiplications in (\ref{eq:D_iter}). The filled area demonstrates the $\pm \sigma$ interval for the distribution of $T_{CG}$ at given parameters: the number of CG iterations can vary from one configuration $\phi^k_x$ to another.}
	\label{fig:full_timing}
\end{figure}

The results of comparison are shown on Fig.~\ref{fig:full_timing}. As expected, the largest speedup is achieved for smaller lattices. In this case the usage of dense linear algebra is advantageous even for sparse initial blocks $D_{2k-1}$ because they become dense too early in the process of Schur iterations. The most important result is that in the strong-coupling phase of the Hubbard model the Schur solver is faster than Conjugate Gradient iterations even for lattices with $N_s = 1000$ sites. Moreover, when the linear algebra routines for sparse matrices are used, the speed-up is by at least a factor of ten, and depends rather weakly on the lattice size. A rough extrapolation of this result suggests that in the strong-coupling phase of the Hubbard model the Schur complement solver will outperform CG iterations for all practically relevant lattice sizes up to at least $N_s \sim 10^4$.

In the weak-coupling phase of the Hubbard model the speed-up is also significant for moderately large lattice sizes, however, the difference with CG is not so large. Again, a rough extrapolation suggests that in this regime the Conjugate Gradient method might become more efficient than the Schur solver at $N_s \sim 10^3 \ldots 10^4$. Likewise, for the Hubbard-Coulomb model in the weak-coupling phase the Schur complement solver outperforms CG iterations only up to $N_s \sim 10^3$.

\begin{figure}[t!]
    \centering
	\includegraphics[width=0.49\textwidth]{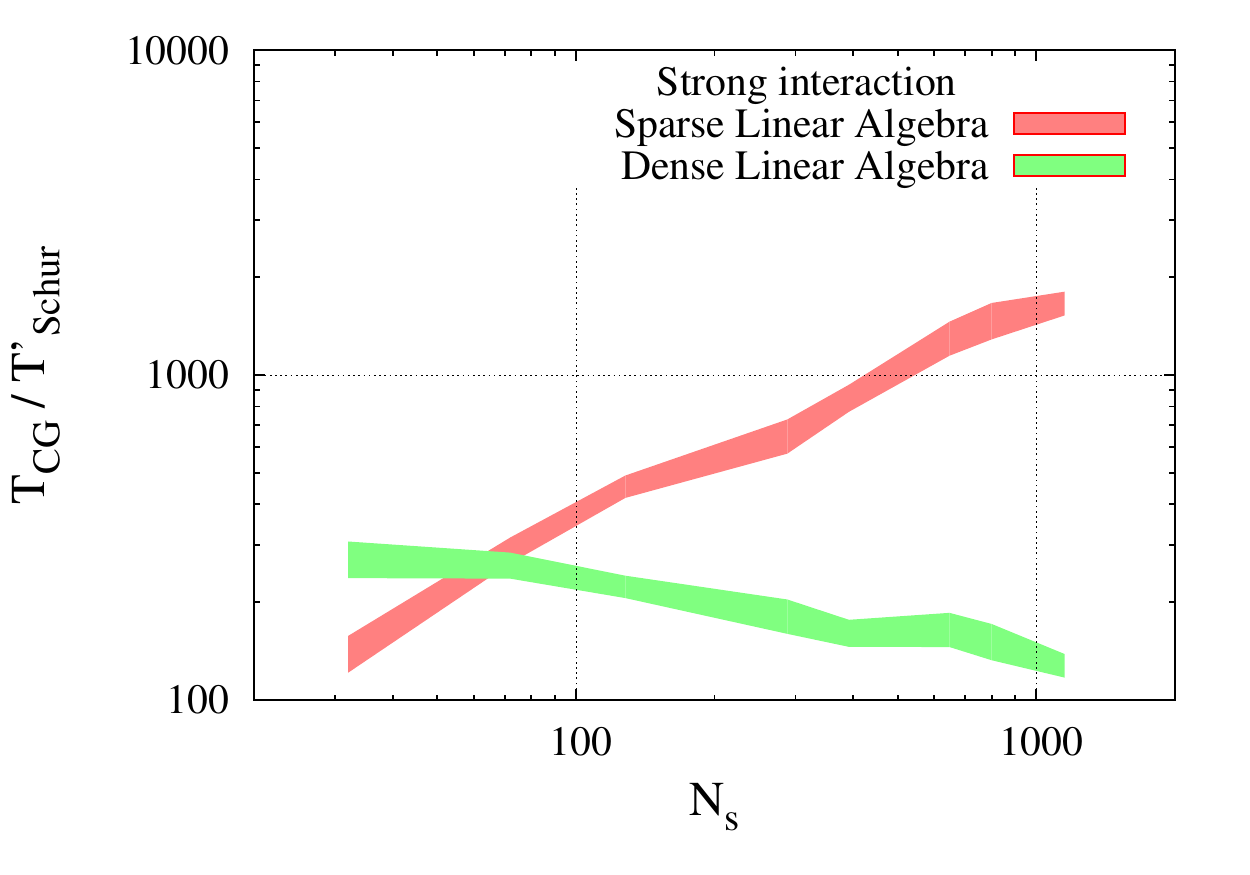}
	\includegraphics[width=0.49\textwidth]{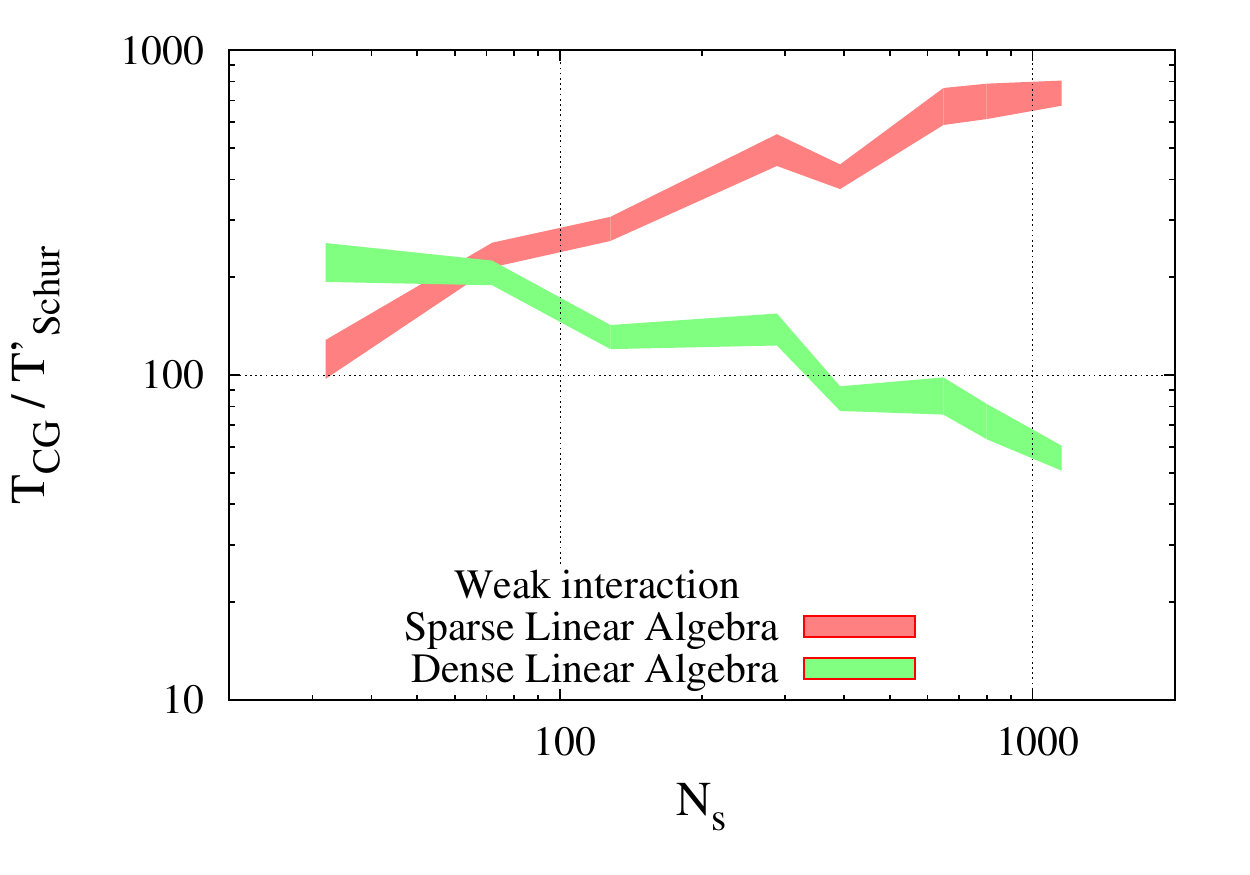}\\
    \includegraphics[width=0.49\textwidth]{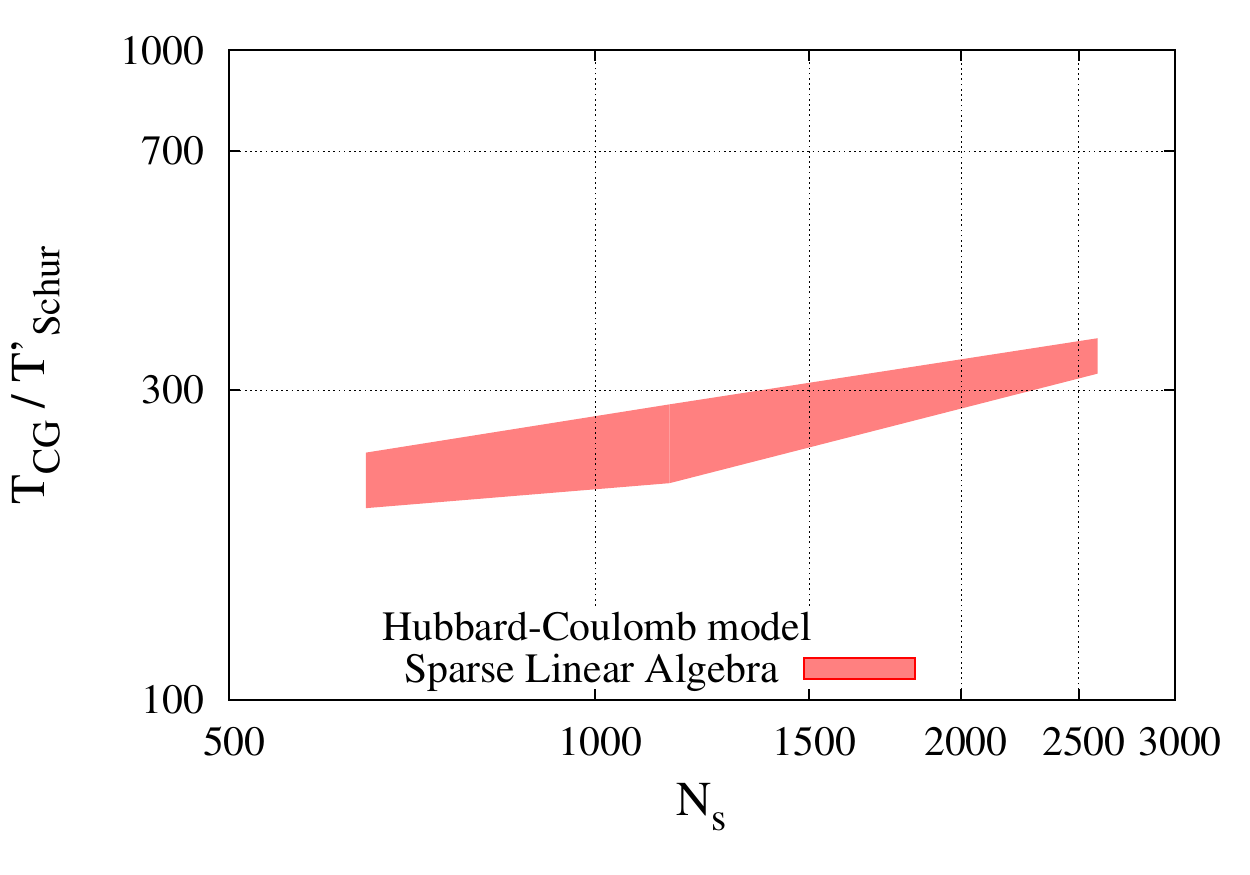}
    \includegraphics[width=0.49\textwidth]{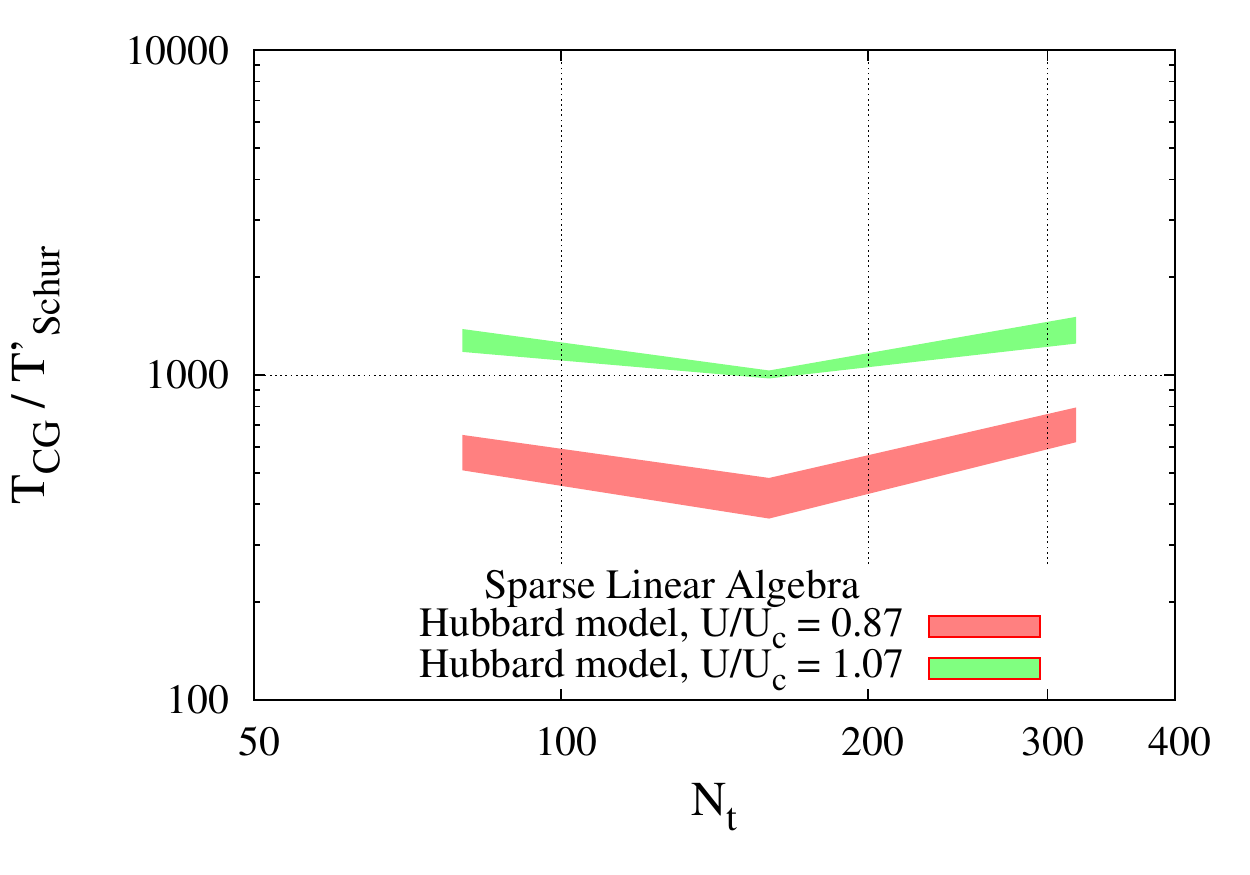}\\
	\caption{Comparison of timings $T_{CG}$ and $T_{Schur}'$ of Conjugate Gradient and Schur complement solver for a single right-hand side in the equations (\ref{eq:lin_sys_real}) excluding the time for matrix-matrix operations within the Schur solver (which can be re-used for multiple right-hand sides). On the top: for the Hubbard model in the strong-coupling regime with $U/U_c = 1.07$ (left plot) and in the weak-coupling regime with $U/U_c = 0.87$ (right plot) at fixed $N_t = 128$ and different $N_s$. Bottom left: for the Hubbard-Coulomb model in the weak coupling regime away from the phase transition ($U/U_c=0.7$) at fixed $N_t = 80$ and different $N_s$. Bottom right: for the Hubbard model at fixed $N_s = 392$ and physical temperature $\beta\kappa=21.6$ and different $N_t$. Similar to the figure \ref{fig:full_timing} we show results for both cases where we use sparse and dense linear algebra, filled areas demonstrate the dispersion of the distribution of $T_{CG}$.}
	\label{fig:multiple_rhs_timing}
\end{figure}

These trends become even more favourable for the Schur solver if we consider the timing $T_{Schur}'$ for multiple solutions of the system (\ref{eq:lin_sys}) with different right-hand sides $Y$. For CG, we still consider the conventional iterations without any speed-up for multiple right-hand sides, while for the Schur solver only matrix-vector operations are involved. These results are illustrated on Fig.~\ref{fig:multiple_rhs_timing}, which has the same layout as Fig.~\ref{fig:full_timing}. We see that in this case the gained speed-up is by a factor of at least $10^2$. Most importantly, the speed-up increases with lattice size $N_s$, thus the Schur solver becomes more and more advantageous for the calculation of observables (\ref{eq:stoch_obser}) with increasing lattice size. 

Again, the gain in performance is more profound in the strongly-correlated regime behind the phase transition. In this case the speedup over CG can reach levels well above $10^{3}$ for large enough lattices. 

\begin{figure}[t!]
    \centering
	\includegraphics[width=0.6\textwidth]{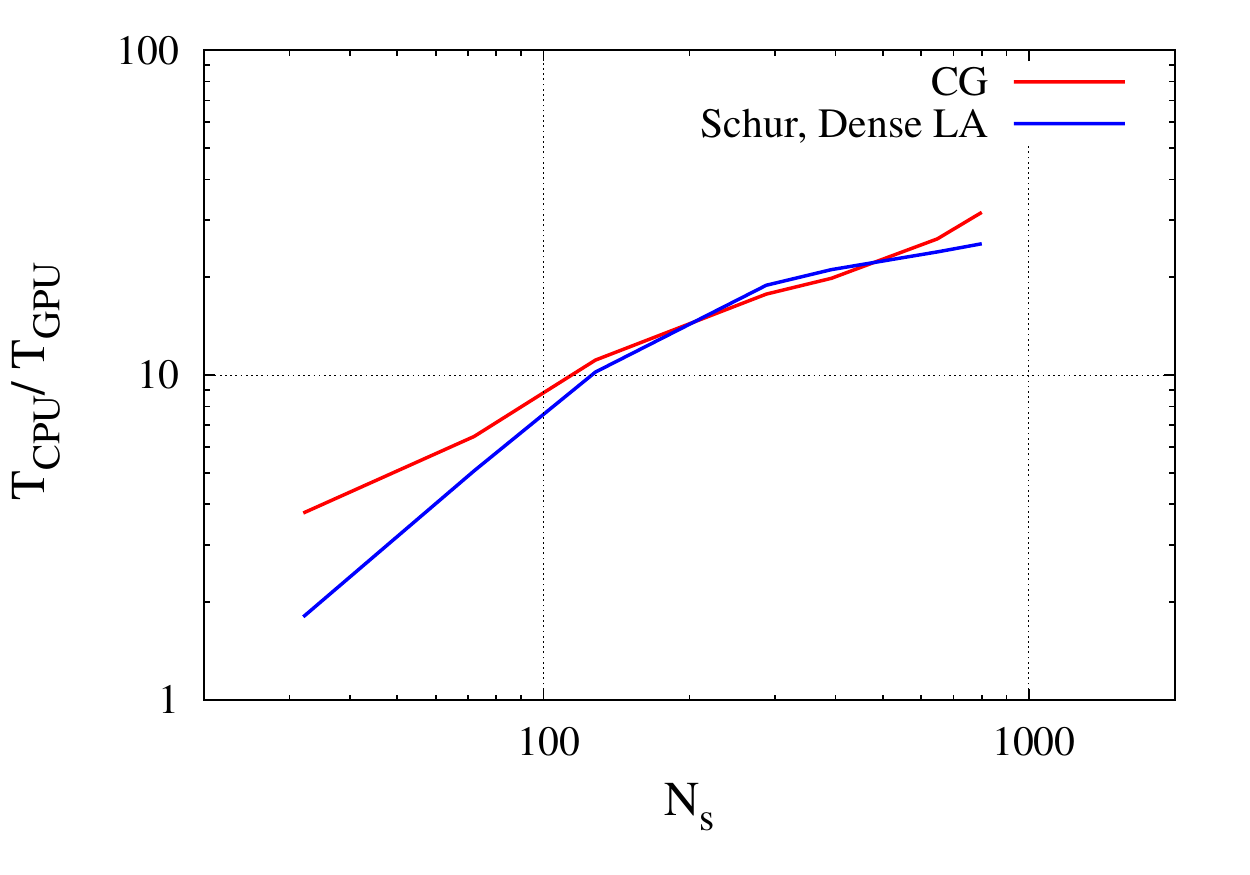}
	\caption{Speedup with parallelization of both solvers on GPU. The comparison is made for the case of the Hubbard model in strongly-correlated regime ($U/U_c = 1.07$), $N_t=128$. For the case of Schur solver,  dense linear algebra routines were used in both (CPU and GPU) cases. The following hardware was used for the tests: Intel Xeon E5645 vs NVIDIA Tesla M2075. }
	\label{fig:gpu_vs_cpu}
\end{figure}

In the next tests we compare the efficiency of the GPU-parallelization of iterative and non-iterative solvers. Timings of single-core CPU and single-GPU implementations of the Schur complement solver are compared. On Fig.~\ref{fig:gpu_vs_cpu} we show the ratios $T_{CPU}/T_{GPU}$ of the computer times required for one solution of system (\ref{eq:lin_sys}) on CPU and on GPU, using either the Schur or the Conjugate Gradient methods. We find that in general CG can be parallelized slightly more efficiently: The GPU version demonstrates larger performance gains for smaller lattices and the saturation seems to appear at larger $N_s$ and at higher levels of speedup. Nevertheless, the existing GPU version of the Schur solver still demonstrates rather efficient parallelization being $\sim 30$ times faster then single-core CPU version, at least for large lattices. 

Finally, we perform several tests of the new solver in full HMC calculations. First of all, we compare timings of two HMC simulations: the first is made using preconditioned Conjugate Gradient and the second uses Schur complement solver to invert the fermion operator. The following setup was used: the Hubbard model on $6\times6\times128$ hexagonal lattice, $U=U_c$ and inverse temperature $\beta$ is equal to 20. We used sparse form of the blocks $D_k$ (\ref{eq:D2k_1_sparse}) for both calculations.

\begin{figure}[t!]
    \centering
	\includegraphics[width=0.6\textwidth]{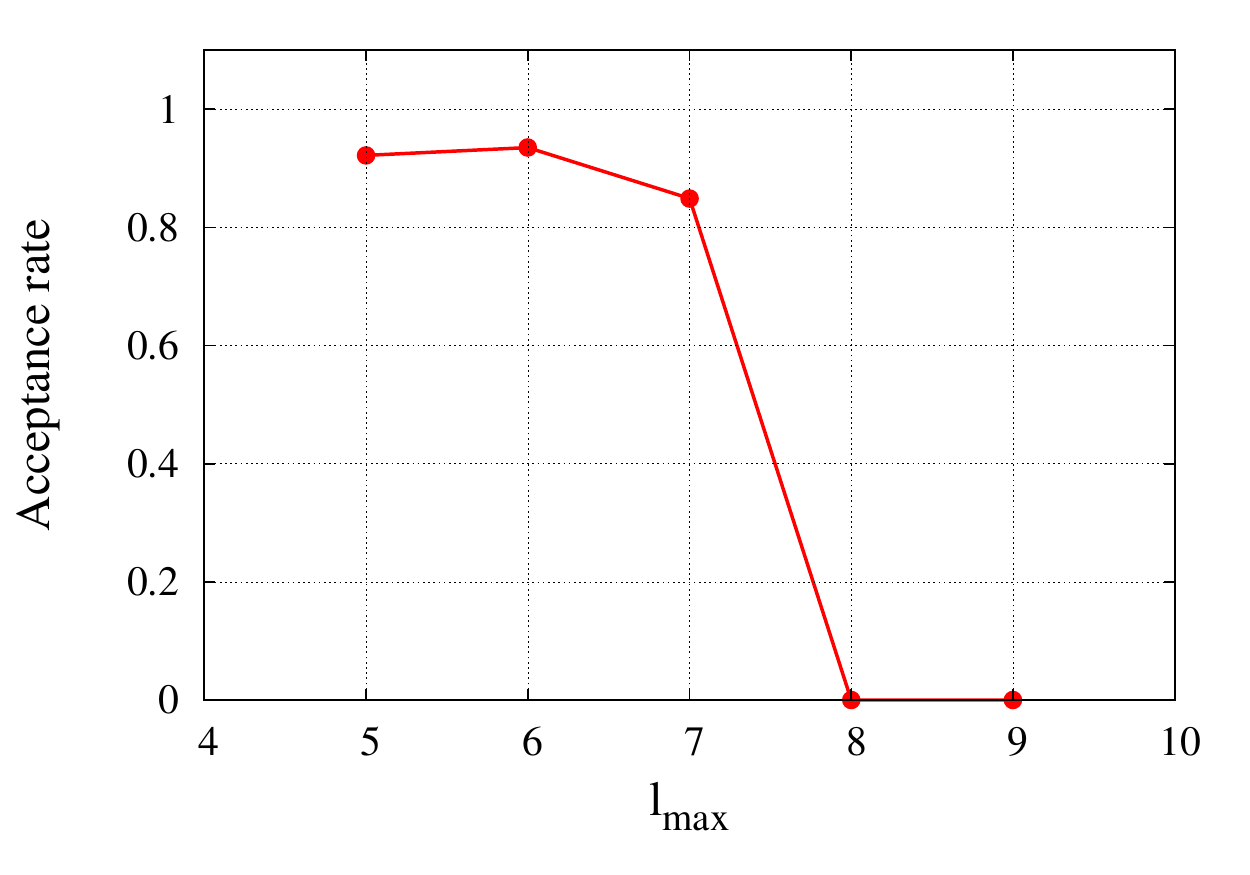}
	\caption{Dependence of acceptance rate in HMC on the number of iterations in Schur solver. Tests were made for Hubbard model on $6\times6\times1024$ hexagonal lattice, $U=U_c$. Metropolis accept-reject step is always made with $l_{max}=5$ to guarantee the stability of final distribution.}
	\label{fig:iterations_acceptance}
\end{figure}

In HMC with pseudofermions, we use the representation of the fermionic determinant (\ref{eq:stoch_det}) and add artificial moment for each continuous bosonic field. As a result, we arrive at the following representation of the partition function: 
\begin{eqnarray}
\mathcal{Z}=\int D \phi D p d\bar Y d Y e^{-H_{eff.}}, \nonumber \\
H_{eff.}= \sum_i \frac{p_i^2}{2} + \frac{\sum_{i,j} \phi_i U^{-1}_{ij} \phi_j}{2 \Delta \tau} + \bar Y (M^\dag M) ^{-1} Y
\label{eq:part_func_HMC}
\end{eqnarray}
Generally, any single update of the field configuration in HMC consists of two stages:
\begin{enumerate}
\item Molecular dynamics according to the Hamiltonian $H_{eff.}$ in (\ref{eq:part_func_HMC}). Since we need to compute derivatives of the last term in the Hamiltonian with respect to the fields $\phi_i$, eq. (\ref{eq:lin_sys}) should be solved at each step of a trajectory. 
\item Metropolis accept-reject step according to the probability distribution $e^{-H_{eff.}}$. At this point the inversion of the fermion operator should be made with higher precision which means reduced $l_{max}$ for Schur solver. It ensures that the probability distribution is still correct even if dynamics on the previous step was not exact. 
\end{enumerate}
The second step in HMC update guarantees the independence of the results on the accuracy of the fermionic force calculation during the trajectory (of course, if the acceptance rate still remains nonzero). Nevertheless, we checked that some important fermionic observables, e.g. the squared spin, are the same within error bars for both runs. As expected, the run with Schur solver demonstrates substantial speedup: it is 29.3 times faster than HMC run with CG. 

In the second test, we checked the effect of increased number of iterations in Schur solver on the HMC performance. Results are shown on Fig. \ref{fig:iterations_acceptance}. We used the same lattice as in the first HMC test, the only difference is enlarged Euclidean time extent: $N_t=1024$ instead of $N_t=128$. The exponential growth of errors with increased number of iterations (see Fig. \ref{fig:accuracy_Ns288_Nt128}) reveals itself in the sharp drop of acceptance rate for $l_{max} \geq 8$. The limiting value can vary of course for different lattice models, but the observed drop in the acceptance rate can be easily used to set up the value of $l_{max}$ in practical calculations.

\section{Conclusions and Summary}
\label{sec:summary}

To conclude, our numerical tests for the strong-coupling regime of the interacting tight-binding models (\ref{eq:full_hamiltonian}) on the hexagonal graphene lattice indicate that the non-iterative Schur complement solver clearly outperforms the Conjugate Gradient method for lattice sizes used in typical numerical simulations, at least for two-dimensional condensed matter systems. Iterative solution is only advantageous in the weak-coupling regime, where the condition number of the fermionic matrix is sufficiently small. An even more dramatic speedup is achieved for multiple right-hand sides of the linear system (\ref{eq:lin_sys}), in which case the Schur complement solver is advantageous both in the strong- and weak-coupling regimes. We have also used the Schur complement solver in several Hybrid Monte-Carlo studies \cite{Ulybyshev:1707.04212,Buividovich:17:1,Buividovich:16:4} which would have taken a much larger time with a Conjugate Gradient solver.

While we have not tested the Schur complement solver in the context of lattice QCD simulations, from eq. (\ref{eq:scaling}) we can readily estimate the number of floating-point operations for some typical lattice parameters. For example, for staggered fermions on a $10^3 \times 64$ lattice in the background of $SU\lr{3}$ gauge fields, the number of floating-point operations in the Schur solver is comparable to approximately 5000 iterations of the Conjugate Gradient algorithm, which is a rather large number for a typical HMC simulation. The situation tends to become worse for larger lattices due to slowing down of LU decomposition. For a $32^3 \times 64$ lattice the Schur solver is equivalent to $3 \times 10^6$ CG iterations. However, the situation changes if we need to solve many right-hand sides of the equation (\ref{eq:lin_sys}) simultaneously. After the construction of all matrix blocks $D^{(l)}_k$ and the final LU decomposition are made, each new solution costs only $N_s^2 + A N_s N_t$ floating-point operations, which is much cheaper than a new solution using CG. This number of operations is equivalent to only about 4 or 5 CG iterations for the $10^3 \times 64$ lattice and about 100 CG iterations for $32^3 \times 64$. Thus the Schur solver can be useful in the calculation of observables. 

Another advantage of the Schur complement solver is that it is insensitive to the positive-definiteness or hermiticity properties of the fermionic matrix. While the Conjugate Gradient method explicitly relies on these properties, iterative methods for non-Hermitian matrices such as GMRes or BiCGStab are also often unstable for matrices which strongly deviate from hermiticity. In this situation, the Schur complement solver might be useful for simulations of finite-density QCD with non-Hermitian (or non-$\gamma_5$-Hermitian) Dirac operators, which are at present limited to rather small lattices. In this context, an important advantage for simulations which use reweighting to deal with the fermionic sign problem is that our solver gives direct access to the phase and the modulus of the fermionic determinant in the process of solving the linear system (this is again a consequence of the fact that the Schur complement solver is a special version of LU decomposition). In contrast to the methods based on the full reduction of the matrix $M$ in (\ref{eq:lin_sys}) to the size $N_s \times N_s$, as in \cite{Hasenfratz:NuclPhysB.317.539,Nakamura:1009.2149}, the Schur complement solver offers better control over the condition number of the reduced matrix and the effect of round-off errors.

Furthermore, the Schur complement solver might be advantageous for simulations of finite-density QCD based on Lefshetz thimble decomposition \cite{Scorzato:12:1} or approximations thereof \cite{Alexandru:1703.02414}, where one can use stochastic estimators to calculate the derivative of the fermionic determinant over gauge fields, as required for the complexified gradient flow.

Finally, Schur complement solver is particularly suitable for working with fermionic matrices with non-sparse spatial blocks $D_i$ in (\ref{eq:M}) - for example, for overlap fermions in the Hamiltonian/canonical formulation \cite{Creutz:01:1}, or for supersymmetric matrix quantum mechanics \cite{Hanada:13:1}.

\section*{Acknowledgments}

M.~U.~was supported by the DFG grant BU 2626/2-1 and by the DFG grant AS120/14-1. P.~B. was supported by the S.~Kowalevskaja award from the Alexander von Humboldt Foundation. K.~K. and N.~K.~were funded by Deutsche
Forschungsgemeinschaft (DFG) Transregional Collaborative Research Centre 55 (SFB/TRR55).



\end{document}